\documentclass[twocolumn,showpacs,preprintnumbers,amsmath,amssymb]{revtex4}

\usepackage{graphicx}
\usepackage{dcolumn}
\usepackage{bm}

\begin{document}


\title{Variant of the Clauser-Horne-Shimony-Holt inequality}
\author{Zeqian Chen}
\email{zqchen@wipm.ac.cn}
\affiliation{%
State Key Laboratory of Magnetic Resonance and Atomic and Molecular
Physics and United Laboratory of Mathematical Physics, Wuhan
Institute of Physics and Mathematics, Chinese Academy of Sciences,
30 West District, Xiao-Hong-Shan, P.O.Box 71010, Wuhan 430071,
China}

\date{\today}

\begin{abstract}
We construct a Bell inequality from the Clauser-Horne-Shimony-Holt
inequality for two qubits that provides a stronger bound on the
correlations of entangled states than allowed by the CHSH
inequality. The argument involved here can be generalized to $n$
qubits. The inequalities obtained are violated by all the
generalized Greenberger-Horne-Zeilinger states of multiqubits.
\end{abstract}

\pacs{03.67.Mn, 03.65.Ud}
\maketitle

Originally, Bell's inequality \cite{Bell} was designed to rule out
various kinds of local hidden variable theories based on Einstein,
Podolsky, and Rosen's (EPR's) notion of local realism \cite{EPR}.
Bell inequalities now serve a dual purpose. On one hand these
inequalities provide a test to distinguish entangled from
non-entangled quantum states. Indeed, experimenters routinely use
violations of Bell inequalities to check whether they have succeeded
in producing entangled states \cite{GN}. On the other hand,
violations of the inequalities are applied to realizing certain
tasks by quantum information, such as building quantum protocols to
decrease communication complexity \cite{BZZ} and making secure
quantum communication \cite{SGA}. Derivations of new and stronger
Bell-type inequalities are thus one of the most important and
challenging subject in quantum theory.

There are extensive earlier works on Bell inequalities \cite{arxiv},
including Clauser-Horne-Shimony-Holt (CHSH) inequality for bipartite
systems \cite{CHSH} and Mermin-Roy-Singh-Ardehali-Belinskii-Klyshko
(MK) inequalities for multi-particle systems \cite{MK}. We refer to
\cite{review} and references therein for more details. In this
article, we construct a Bell inequality from the
Clauser-Horne-Shimony-Holt (CHSH) inequality for two qubits that
provides a stronger bound on the correlations of entangled states
than allowed by the CHSH inequality. Thus, a sharper experimental
criterium is obtained for the distinction between entangled and
non-entangled states. The argument involved can be generalized to
$n$ qubits. The inequalities obtained can be used to detect all the
generalized Greenberger-Horne-Zeilinger (GHZ) states of multiqubits.

Let us consider a system of two qubits labelled by $1$ and $2.$ Let
$A,A'$ denote spin observables on the first qubit, and $B,B'$ on the
second. For $A^{(\prime)}= \vec{a}^{(\prime)} \cdot \vec{\sigma}_1$
and $B^{(\prime)} = \vec{b}^{(\prime)} \cdot \vec{\sigma}_2,$ we
write$$(A,A') = (\vec{a}, \vec{a}' ), A \times A' = ( \vec{a} \times
\vec{a}') \cdot \vec{\sigma}_1,$$and similarly,$(B,B')$ and $B
\times B'.$ Here $\vec{\sigma}_1$ and $\vec{\sigma}_2$ are the Pauli
matrices for qubits $1$ and $2,$ respectively; the norms of real
vectors $\vec{a}^{(\prime)}, \vec{b}^{(\prime)}$ are equal to $1.$
If $A,A'$ ($B,B'$) are orthogonal, denoted by $A \perp A',$ then
$A'' = A \times A'$ ($B'' = B \times B'$) is a spin observable.

We write $AB,$ etc., as shorthand for $A \otimes B$ and $\langle A B
\rangle_{\varrho} : = \mathrm{Tr} \varrho A \otimes B;$ $ \langle A
B \rangle_{\psi} = \langle \psi | A  B | \psi \rangle$ for the
expectations of $A B$ in the mixed state $\varrho$ or pure state
$\psi.$ Then, the CHSH inequality holds:\begin{equation}\label{1}|
\langle AB + AB' + A' B - A' B' \rangle_{\varrho} | \leq
2,\end{equation}for all such observables and all non-entangled
states, i.e., states of the form $\varrho = \varrho_1 \otimes
\varrho_2$, or convex mixtures of such states. The maximal violation
of the Eq.\eqref{1} for entangled states is $2 \sqrt{2},$ which is
attained only by Bell states $$| \phi^{\pm} \rangle = \frac{1}{2}
\left ( |\uparrow \uparrow \rangle \pm |\downarrow \downarrow
\rangle \right ), | \psi^{\pm} \rangle = \frac{1}{2} \left (
|\uparrow \downarrow \rangle \pm |\downarrow \uparrow \rangle \right
)$$ and the states obtained from them by local unitary
transformations \cite{C80}, i.e., the states of form $U_1 U_2 |
\phi^{+} \rangle,$ where $U_1, U_2$ are unitary transformations on
qubit 1 and qubit 2 respectively.

We present a variant of the CHSH as follows: For all orthogonal spin
observables $A,A'$ on the first qubit with $A'' = A \times A'$ and
$B,B'$ on the second with $B'' = B \times
B'$,\begin{equation}\label{2}| \langle AB + AB' + A' B - A' B' + 2
A'' B'' \rangle_{\varrho} | \leq 2,\end{equation} for all
non-entangled states $\varrho,$ and\begin{equation}\label{3}|
\langle AB + AB' + A' B - A' B' + 2 A'' B'' \rangle_{\varrho} | \leq
2 (1 + \sqrt{2}),\end{equation} for all entangled states $\varrho,$
respectively.

Moreover, every entangled pure state $\psi$ can violate Eq.(2),
i.e., there are orthogonal spin observables $A,A'$ on the first
qubit and orthogonal $B,B'$ on the second such
that\begin{equation}\label{4} \langle AB + AB' + A' B - A' B' + 2
A'' B'' \rangle_{\psi}
> 2.\end{equation}Thus, Gisin's theorem \cite{G} for the CHSH inequality
Eq.\eqref{1} holds for Eq.\eqref{2} too.

The equality in Eq.\eqref{3} is attained only by the Bell states and
the states obtained from them by local unitary transformations.
Then, the maximal violation of the Eq.\eqref{2} for entangled states
is $2 (1 + \sqrt{2}),$ which is a stronger bound on the correlations
of entangled states than $2 \sqrt{2}$ allowed by the CHSH inequality
Eq.\eqref{1}.

To prove Eqs.\eqref{2},\eqref{3} and \eqref{4}, we note first that
these inequalities are all invariant under local unitary
transformations, and then we can assume that $A = \sigma^1_x, A' =
\sigma^1_y$ and so $A''=\sigma^1_z,$ as well as $B = \sigma^2_x, B'
= \sigma^2_y$ and so $B''=\sigma^2_z.$ We
have\begin{equation}\label{5}\begin{split} \mathcal{B}_2 & =
\sigma^1_x \otimes \frac{1}{2} ( \sigma^2_x + \sigma^2_y ) +
\sigma^1_y \otimes \frac{1}{2} (\sigma^2_x - \sigma^2_y) \\
& = \frac{1}{2} (\sigma^1_x \sigma^2_x + \sigma^1_x \sigma^2_y +
\sigma^1_y \sigma^2_x - \sigma^1_y \sigma^2_y )\\
& = \sqrt{2} \left ( | \varphi^+ \rangle \langle \varphi^+ | -
|\varphi^-\rangle \langle \varphi^- | \right
),\end{split}\end{equation}where $| \varphi^{\pm} \rangle =
\frac{1}{\sqrt{2}}\left ( |00 \rangle \pm e^{i\pi /4} |11 \rangle
\right )$ with $\sigma_z |0 \rangle = |0 \rangle$ and $\sigma_z |1
\rangle = -| 1 \rangle .$ Since$$\mathcal{B}^2_2 = 1 + \sigma^1_z
\sigma^2_z,$$Eqs.\eqref{2} and \eqref{3} are equivalent
to\begin{equation}\label{6}\langle \mathcal{B}_2 + \mathcal{B}^2_2
\rangle \leq 2 \end{equation}and\begin{equation}\label{7}\langle
\mathcal{B}_2 + \mathcal{B}^2_2 \rangle \leq 2 +
\sqrt{2},\end{equation}respectively. It is concluded from
Eq.\eqref{5} that $\| \mathcal{B}_2 \| \leq \sqrt{2}$ and hence
Eq.\eqref{7} holds. It remains to prove Eq.\eqref{6}.

To this end, we note that $\mathcal{B}_2 + \mathcal{B}^2_2 =
\sqrt{2} ( e^{i \pi /4} |11 \rangle \langle 00 | + e^{- i \pi /4}|00
\rangle \langle 11 |) + 2 ( |00 \rangle \langle 00 | + |11 \rangle
\langle 11 |).$ Then, for every product state $| \psi \rangle = |
\psi_1 \rangle | \psi_2 \rangle$ of $2$ qubits, one
has\begin{equation*}\begin{split}\langle \mathcal{B}_2 +
\mathcal{B}^2_2 \rangle_{\psi} & = \sqrt{2} (e^{i\pi /4}
\beta_1\beta_2 \alpha^*_1\alpha^*_2 + e^{- i\pi /4} \alpha_1\alpha_2
\beta^*_1\beta^*_2 )\\
&~~ + 2 (|\alpha_1 \alpha_2 |^2 + |\beta_1 \beta_2 |^2)\\
&\leq \sqrt{2} (| \alpha_1 \alpha_2 | + |\beta_1 \beta_2 |)^2\\
&~~+ (2- \sqrt{2}) (|\alpha_1 \alpha_2 |^2 + |\beta_1 \beta_2 |^2),
\end{split}\end{equation*}where
$\alpha_j = \langle \psi_j | 0 \rangle$ and $\beta_j = \langle
\psi_j | 1 \rangle,$ $j=1,2.$ Using $\max (x \sin \phi + y \cos \phi
) = \sqrt{ x^2 + y^2},$ we get $(| \alpha_1 \alpha_2 | + |\beta_1
\beta_2 |)^2 \leq 1.$ This concludes Eq.\eqref{6}.

By the Schmidt decomposition theorem, every pure state $| \psi
\rangle$ of two qubits is of the form$$| \psi (\theta) \rangle =
\cos \theta |00 \rangle + e^{i \pi /4}\sin \theta |11 \rangle$$for
$0 \leq \theta \leq \pi /4,$ under a local unitary transformation.
Then, $| \psi \rangle$ is entangled if and only if $\theta >0.$ It
is easy to check that$$\langle \psi (\theta) | \mathcal{B}_2 +
\mathcal{B}^2_2 | \psi (\theta) \rangle = \sqrt{2}\sin 2 \theta +
2.$$Consequently,\begin{equation*}\begin{split}\langle \sigma^1_x
\sigma^2_x &+ \sigma^1_x \sigma^2_y + \sigma^1_y \sigma^2_x -
\sigma^1_y \sigma^2_y - 2 \sigma^1_z \sigma^2_z \rangle_{\psi
(\theta)}\\
& = \sqrt{2}\sin 2 \theta + 2 > 2\end{split}\end{equation*}whenever
$\theta >0.$ This shows that the inequality \eqref{4} holds.

The extension of the argument involved above for $n$-qubits is
straightforward. Note that the MK Bell operators $\mathcal{B}_n$ of
$n$ qubits ($n > 2$) are defined recursively:
\begin{equation}\label{8}{\cal B}_n = {\cal B}_{n-1} \otimes
\frac{1}{2}(\sigma^n_x + \sigma^n_y) + {\cal B}'_{n-1} \otimes
\frac{1}{2}( \sigma^n_x - \sigma^n_y ),\end{equation}where ${\cal
B}'_n$ denotes the same expression ${\cal B}_n$ but with all the
$\sigma_x$ and $\sigma_y$ exchanged. The $n$-qubit inequalities for
the extension of Eqs.\eqref{2} and \eqref{3} are as
follows:\begin{equation}\label{9}\langle {\cal B}_n + \sum_{i<j}
\sigma^i_z \sigma^j_z + \sum_{i<j<k<l} \sigma^i_z \sigma^j_z
\sigma^k_z \sigma^l_z +\cdots \rangle_{\varrho} \leq 2^{n-1}
-1,\end{equation}for all non-entangled states $\varrho,$ and
\begin{equation}\label{10}\begin{split}\langle
{\cal B}_n + \sum_{i<j} \sigma^i_z \sigma^j_z & + \sum_{i<j<k<l}
\sigma^i_z \sigma^j_z \sigma^k_z \sigma^l_z +\cdots
\rangle_{\varrho}\\
& \leq 2^{(n-1)/2} + 2^{n-1} -1,\end{split}\end{equation} for all
entangled states $\varrho,$ respectively.

Furthermore, the inequality \eqref{9} can be violated by all the
generalized Greenberger-Horne-Zeilinger
states\begin{equation}\label{11}| \Psi (\theta) \rangle = \cos
\theta |0^n \rangle + e^{i \pi /4} \sin \theta | 1^n \rangle,
\end{equation}with $0 < \theta \leq \pi /4.$ The GHZ state \cite{GHZ}
is for $\theta = \pi /4.$ Here, we adopt the notation $|0^n \rangle
= |0 \cdots 0 \rangle$ and $|1^n \rangle = |1 \cdots 1 \rangle.$

The proofs of Eqs.\eqref{9},\eqref{10} and \eqref{11} are similar to
the case of two qubits. Indeed,
since\begin{equation*}\mathcal{B}^2_n = 1 + \sum_{i<j} \sigma^i_z
\sigma^j_z + \sum_{i<j<k<l} \sigma^i_z \sigma^j_z \sigma^k_z
\sigma^l_z +\cdots\end{equation*}(e.g., see \cite{SG}), the
inequalities \eqref{9} and \eqref{10} are equivalent
to\begin{equation}\label{12}\langle \mathcal{B}_n + \mathcal{B}^2_n
\rangle \leq 2^{n-1}
\end{equation}and\begin{equation}\label{13}\langle \mathcal{B}_2 +
\mathcal{B}^2_2 \rangle \leq 2^{(n-1)/2} +
2^{n-1},\end{equation}respectively. Recall that the MK Bell operator
$\mathcal{B}_n$ has the following spectral decomposition
\cite{SG}:\begin{equation}\label{14}\mathcal{B}_n = 2^{(n-1)/2}\left
( |\textrm{GHZ}_+\rangle \langle \textrm{GHZ}_+ | -
|\textrm{GHZ}_-\rangle \langle \textrm{GHZ}_- | \right
),\end{equation}where $|\textrm{GHZ}_{\pm} \rangle =
\frac{1}{\sqrt{2}}\left ( |0^n \rangle \pm e^{i \pi /4} |1^n \rangle
\right )$ are the GHZ states \cite{GHZ}. Then, $\| \mathcal{B}_n \|
\leq 2^{(n-1)/2}$ and so the inequality \eqref{13} holds true.

To prove the inequality \eqref{12}, note that by \eqref{14}
\begin{equation}\label{15}\mathcal{B}_n = 2^{(n-1)/2} (e^{-i\pi/4}
|0^n \rangle \langle 1^n | + e^{i\pi/4}|1^n \rangle \langle 0^n
|)\end{equation}and\begin{equation}\label{16}\mathcal{B}^2_n =
2^{n-1} ( |0^n \rangle \langle 0^n | + |1^n \rangle \langle 1^n
|).\end{equation}Then, for every product state $| \psi \rangle = |
\psi_1 \rangle\cdot\cdot\cdot| \psi_n \rangle$ of $n$ qubits, one
has\begin{equation*}\begin{split}\langle \mathcal{B}_n +
\mathcal{B}^2_n \rangle & = 2^{(n-1)/2} \Big ( e^{-i\pi /4}
\prod^n_{j=1} \alpha_j \beta^*_j + e^{i\pi /4} \prod^n_{j=1}
\alpha^*_j \beta_j \Big )\\
&~~ + 2^{n-1} \Big ( \prod^n_{j=1} |\alpha_j|^2 +
\prod^n_{j=1} |\beta_j|^2 \Big )\\
& \leq 2^{(n-1)/2} \Big ( \prod^n_{j=1} |\alpha_j| +
\prod^n_{j=1} |\beta_j| \Big )^2\\
&~~ + (2^{n-1}- 2^{(n-1)/2} )\Big (\prod^n_{j=1} |\alpha_j|^2 +
\prod^n_{j=1} |\beta_j|^2 \Big ),\end{split}\end{equation*}where
$\alpha_j = \langle \psi_j | 0 \rangle$ and $\beta_j = \langle
\psi_j | 1 \rangle,$ $j=1,\ldots, n.$ Using $\max (x \sin \phi + y
\cos \phi ) = \sqrt{ x^2 + y^2},$ we get $(\prod^n_{j=1} |\alpha_j|
+ \prod^n_{j=1} |\beta_j|)^2 \leq 1$ for $n \geq 2.$ This concludes
that Eq.\eqref{12} holds for all product states. Since a separable
state is a convex combination of product states, it is concluded
that Eq.\eqref{9} holds for all separable quantum states of $n \geq
2$ qubits.

Moreover, from Eqs.\eqref{15} and \eqref{16} it is concluded that
for the generalized GHZ states $| \Psi (\theta) \rangle$ of
Eq.\eqref{11},\begin{equation}\label{17}\langle \Psi (\theta) |
\mathcal{B}_n + \mathcal{B}^2_n | \Psi (\theta) \rangle =
2^{(n-1)/2} \sin 2 \theta + 2^{n-1}.\end{equation}Thus, all
generalized GHZ entangled states violate Eq.\eqref{9} and, the
bounds of Eqs.\eqref{9} and \eqref{10} are both tight.

The inequalities presented here provide experimentally feasible
means of testing whether quantum states are entangled and the
violation of the inequality \eqref{9} is a sufficient condition for
all the generalized GHZ entangled states. These conditions may be
useful, since they are sum of expectations of local spin observables
which recent experiments can perform. We would like to point out
that although the inequalities \eqref{9} and \eqref{12} are
mathematically equivalent, the inequality \eqref{12} cannot be used
to detect entangled states because the operator $\mathcal{B}^2$
cannot be measured locally. A different explanation for the
inequality \eqref{12} was presented in \cite{Chen} and commented by
\cite{Seevinck}. We will discuss that issue in another paper.

A final remark concerns the relation between testing the
entanglement of quantum states and testing quantum mechanics against
local hidden variable (LHV) theories. As already demonstrated by
Werner \cite{Werner}, testing for entanglement within quantum
theory, and testing quantum mechanics against LHV theories are not
equivalent. Indeed, as shown in \cite{Roy}, the physical origins of
EPR's local realism and quantum entanglement are different. The
bound of inequality \eqref{2} for LHV is clearly $4$ which is larger
than the separability bound of quantum states, but smaller than $2
(1 + \sqrt{2})$ and hence can also be used to testing quantum
mechanics against LHV theories. Similar results also hold for the
inequality \eqref{9}.

In summary, we have presented a family of Bell-type inequalities,
which are constructed directly from the ``standard" Bell
inequalities involving two dichotomic observables per site. It is
shown that the inequalities provide a stronger bound on the
correlations of entangled states than allowed by the CHSH inequality
and are violated by all the generalized GHZ entangled states of
multiqubits.

The author is grateful to Dr.Qingyu Cai for helpful discussions.
This work was partially supported by the National Science Foundation
of China under Grant No.10571176 and 10775175.


\end{document}